\documentclass[conference]{IEEEtran}
\IEEEoverridecommandlockouts
\usepackage{cite}
\usepackage{amsmath,amssymb,amsfonts}
\usepackage{algorithmic}
\usepackage{graphicx}
\usepackage{textcomp}
\usepackage{xcolor}
\def\BibTeX{{\rm B\kern-.05em{\sc i\kern-.025em b}\kern-.08em
    T\kern-.1667em\lower.7ex\hbox{E}\kern-.125emX}}
\begin{document}

\title{Dispersion of Surface Waves above Time-Varying Reactive Boundaries\\
\thanks{This work was supported by the Academy of Finland under grant 330260.}
}

\author{\IEEEauthorblockN{Xuchen~Wang, Mohammad~S.~Mirmoosa, and Sergei~A.~Tretyakov}
\IEEEauthorblockA{\textit{Department of Electronics and Nanoengineering, Aalto University}, Espoo, Finland\\
xuchen.wang@aalto.fi}
} 

\maketitle

\begin{abstract}
In this presentation, we analytically derive the dispersion equation for surface waves traveling along reactive boundaries which are periodically modulated in time. In addition, we show numerical results for the dispersion curves and importantly uncover that time-varying boundaries generate band gaps that can be controlled by engineering the modulation spectrum. Furthermore, we also point out an interesting effect of field amplification related to the existence of such band gaps for surface waves. The effect of amplification does not require the synchronization of signal and pumping waves. This unique property is very promising to be applied in surface-wave communications from microwave to optical frequencies.
\end{abstract}

\begin{IEEEkeywords}
Surface waves, temporal modulation, metasurfaces, dispersion relation.
\end{IEEEkeywords}

\section{Introduction}

It is well known that volumetric photonic crystals have the efficacy of controlling  propagating  waves, e.g.~\cite{joannopoulos2008molding}, both in three and two dimensions, e.g. \cite{Kraus}. 
Recently, in parallel with thorough and general investigations of time-varying electromagnetic systems (see, e.g., Refs.~\cite{XUCHENPRAPPLIED, SAJJADarXiv, groupEnergy}), three-dimensional (volumetric) temporal photonic crystals whose material properties are uniform in space but changing in time have attracted significant attention~\cite{zurita2009reflection}. This is due to the intriguing effects that they have on propagating plane waves. 

It appears that it is fundamentally important to explore properties of temporal metasurfaces which are 2D material sheets or boundaries with time-varying properties. In this case, the waves are surface waves, bound to the sheet or boundary. The eigenmode problem for waves along spatially uniform but time-modulated boundaries is one of the important canonical problems in electromagnetics of time-varying structures.

In this work, we solve this problem and discuss dispersion properties of surface waves over time-modulated reactive boundaries. As a simple canonical case, we consider a planar infinite boundary that is modeled by a surface capacitance which is spatially uniform over the surface and temporally modulated by an arbitrary periodical function. As a particular realization, one can consider, for example, a high-impedance surface of the type introduced by D. Sievenpiper \cite{sievenpiper1999high}, where the capacitance between patches is modulated using varactors.

In addition to derivation of the dispersion relation for surface waves over such time-varying boundary, we give numerical examples of dispersion plots. Also, we explain the conditions for appearing stop bands for propagation constants and reveal phenomena of exponential field growth, reflection amplification, and radiation of space waves from those time-modulated boundaries.



\section{Theory}

\begin{figure}
\includegraphics[width=.95\linewidth]{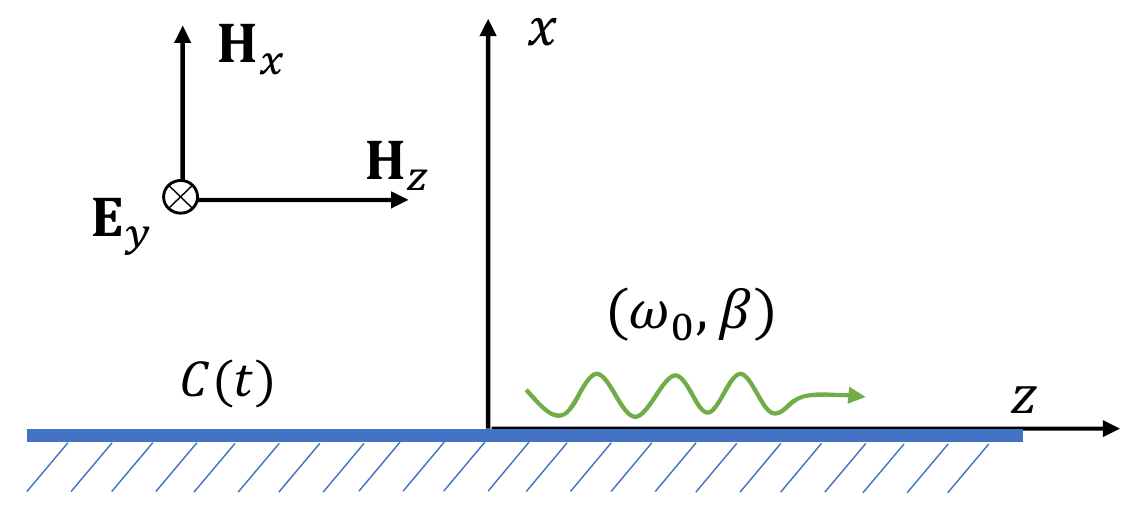}
\caption{Geometry of the problem: A TE-polarized surface wave over a time-modulated capacitive boundary. }
\label{fig:geom}
\end{figure}

Let us consider a spatially uniform and time-varying reactive boundary. Here, as an example, we assume a time-varying capacitive one which is represented by $C(t)$. As is well known, such a boundary supports surface waves which have the transverse-electric (TE) polarization with respect to the propagation direction. In other words, in free space, the electric field is perpendicular to the propagation plane, as shown in Fig.~\ref{fig:geom}. For a stationary boundary, the dispersion equation defines a relation between the frequency $\omega$ and the propagation constant along the surface $\beta$. Since the capacitive boundary periodically changes in time, the eigenmode $\beta$ will contain components at frequencies $\omega_n=\omega+n\omega_{\rm{M}}$, $n=0,\pm 1,\pm 2,\dots$. Here, $\omega_{\rm{M}}$ is the fundamental modulation frequency. The electric field is expressed as 
\begin{equation}
\mathbf{E}=\sum_{n=-\infty}^{+\infty}E_n\exp(j\omega_nt)\mathbf{a}_y,
\end{equation}
in which 
\begin{equation}
E_n=A_n\exp(j\beta z)\exp(-\alpha_nx).
\end{equation}
Here, $A_n$ is the amplitude of the wave corresponding to each frequency harmonic,  and $\alpha_n$ denotes the attenuation constant along the normal direction, for each harmonic.  The plane-wave dispersion equation for free space above the boundary sets the following relation for each harmonic: 
\begin{equation}
\beta^2=\alpha_n^2+\omega_n^2\epsilon_0\mu_0 .
\label{eq:betaalpha}
\end{equation}
It is worth mentioning that since we investigate surface waves, $\alpha_n$ must be a real value.

Similarly to the electric field, the tangential component of the magnetic field which is directed along the $z$-axis is given by 
\begin{equation}
\mathbf{H}_{\rm{t}}=\sum_{n=-\infty}^{+\infty}H_n\exp(j\omega_nt)\mathbf{a}_z,
\end{equation}
where
\begin{equation}
H_n=B_n\exp(j\beta z)\exp(-\alpha_nx).
\end{equation}
The Maxwell  equations relate the amplitudes of the electric field and the tangential component of the magnetic field to each other. By applying Eq.~\eqref{eq:betaalpha} and doing some algebraic manipulations, we obtain a matrix relation  $\mathbf{M}\cdot \mathbf{A}=\mathbf{B}$. Here, $\mathbf{M}$ is a matrix that has $2N+1$ rows and columns, and it is a function of $\beta$ and $\omega_n$. The matrices $\mathbf{A}$ and $\mathbf{B}$ have only one column and $2N+1$ rows, and they are representing the amplitudes.  

In analogy with the circuit theory, where we explicitly express the relation between the electric current flowing through the time-varying capacitance $i(t)$ and the voltage over it $v(t)$ as 
\begin{equation}
\int i(t)dt=C(t)v(t),
\end{equation}
we simply write the relation between the tangential components of the electric and magnetic fields. In fact, this is the boundary condition in the dynamic scenario. By imposing the boundary condition, we obtain another matrix equation as $\mathbf{Y}\cdot \mathbf{A}=\mathbf{B}$. The matrix $\mathbf{Y}$ is a function of $\omega_n$ and the Fourier coefficients of the periodic function $C(t)$ which is expressed by the Fourier series in the exponential form. Similar modeling method has been used in our recent paper \cite{wang2020nonreciprocity}.

Now, we have two matrix equations $\mathbf{M}\cdot \mathbf{A}=\mathbf{B}$ and\break $\mathbf{Y}\cdot \mathbf{A}=\mathbf{B}$. Therefore, we conclude that $\big[\mathbf{Y}-\mathbf{M}\big]\cdot\mathbf{A}=0$.  The determinant of the whole matrix in the square brackets must be zero in order to allow nonzero solutions for the electric field. Consequently,  relation 
\begin{equation}
\det\Big[\mathbf{Y}-\mathbf{M}\Big]=0 \label{eq: dispersion}
\end{equation}
determines the dispersion of the surface waves above  time-varying capacitive boundaries. In the following, we give some particular examples and numerically investigate the dispersion curves.


\section{Numerical Examples and Discussion}

\begin{figure}[tb]
	\centering
	\includegraphics[width=0.95\linewidth]{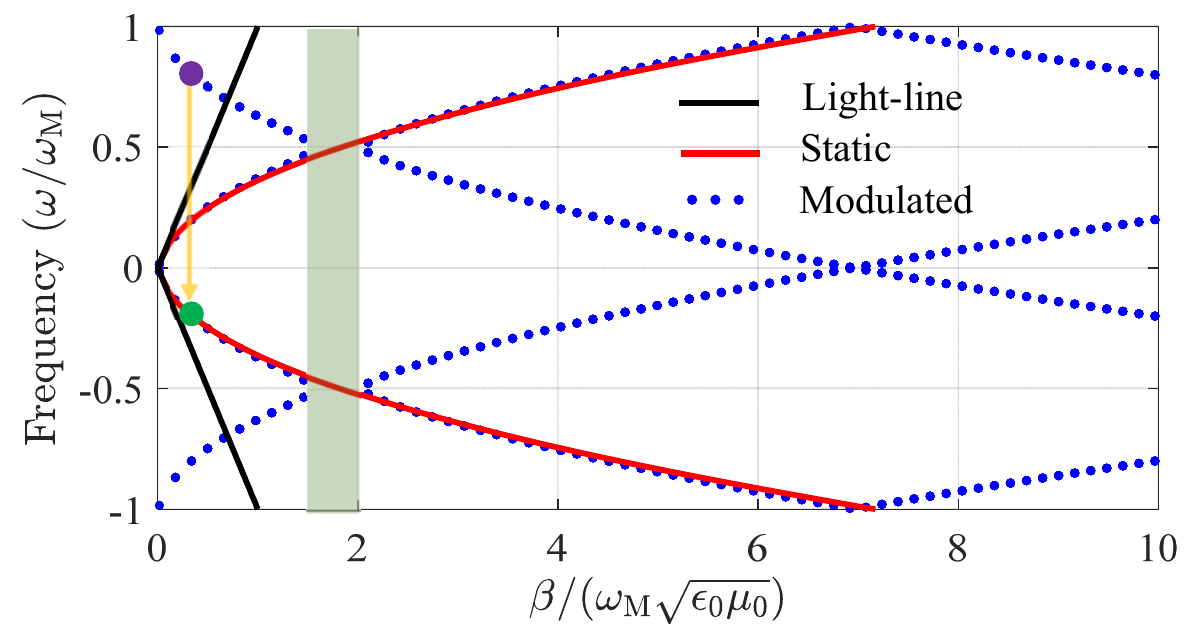}
	\caption{Dispersion plot for modulation with one harmonic tune. The figure is plotted for $C_0=1$~pF and $\omega_{\rm M}=3$~GHz. 
	} 
	\label{fig: dispersion_one_tunes}
\end{figure}
\begin{figure}[!h]
	\centering
	\includegraphics[width=0.95\linewidth]{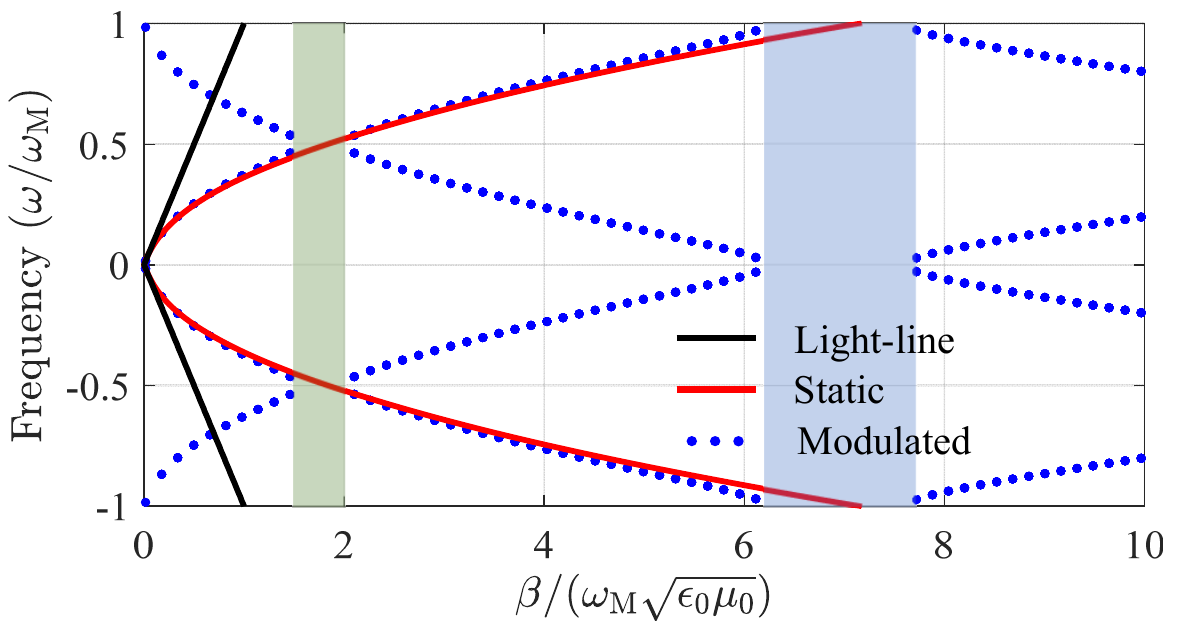}
	\caption{Dispersion plot for modulation with two harmonic tunes:  $C(t)=C_0[1+0.3\cos(\omega_{\rm M}t)+0.2\cos(2\omega_{\rm M}t)]$. 
	} 
	\label{fig: dispersion_two_tunes}
\end{figure}

First, we consider a capacitive boundary that is modulated in time harmonically, assuming, as an example,  $C(t)=C_0[1+0.3\cos(\omega_{\rm M}t)]$. 
To obtain the dispersion diagram, we specify the value of $\beta$, and solve the dispersion equation Eq.~(\ref{eq: dispersion})  for the corresponding  eigenfrequencies. The corresponding dispersion curves are shown in  Fig.~\ref{fig: dispersion_one_tunes}. One  can see  that for a fixed value of the propagation constant $\beta$, there are many solutions for the eigenfrequencies $\omega$, meaning that an eigenmode contains components at many frequencies. This property means that we can excite the mode by external sources at  many possible frequencies. 
The excitation frequency can even be above the light line (purple dot) which excites higher-order frequency harmonics below the light line (green dot).  This indicates that, in time-varying structures, it is possible to launch surface waves with an incident plane wave, as is also reported in a recent work \cite{galiffi2020wood}.
Most importantly, temporal modulation opens up a band gap in $k$-space, which is a dual phenomenon of spatial periodic structures where the band gap is at the frequency axis. Similar properties of bulk media have been reported in \cite{zurita2009reflection,lustig2018topological}.
Increasing the modulation depth, one  can widen the band gap. Interestingly, the number of band gaps corresponds to the number of modulation tones. Figure~\ref{fig: dispersion_two_tunes} shows that by  adding a second modulation tone, a second band gap opens up. Positions and widths of the gaps can be tuned by varying the modulation spectrum. These band gaps provide great possibilities to control the propagation of surface waves on a metasurface plane. 

\begin{figure}[!h]
\includegraphics[width=.95\linewidth]{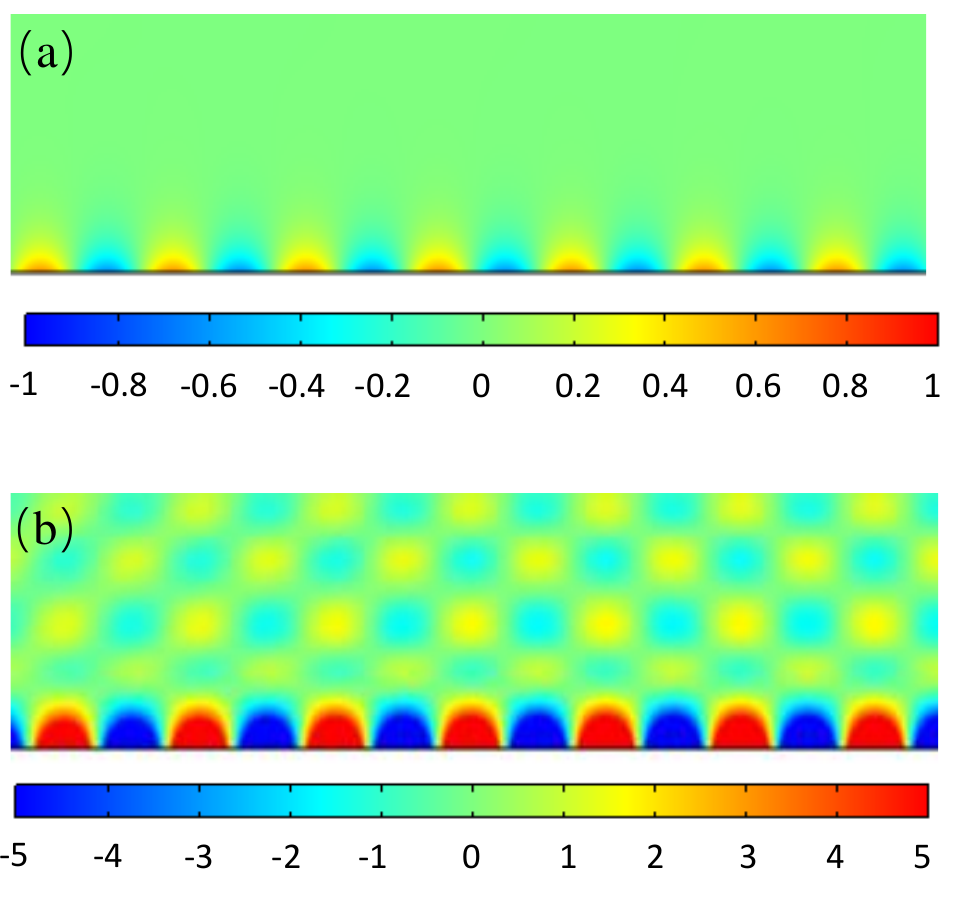}
\caption{(a) Surface wave on a time-invariant boundary. (b) Amplified surface wave on a time-varying boundary ($\Delta t=5T_0$).  }
\label{fig:amplification}
\end{figure}

Next, we examine the wave behavior when the excited surface modes are inside a band gap. For a specified wavenumber $\beta$ in the band gap, the solved eigenfrequencies are complex numbers $\omega=\omega^\prime\pm j\omega^{\prime\prime}$, meaning that  the wave can be exponentially attenuated or amplified in time. Next, we use COMSOL to numerically simulate the time-varying structure in this regime. 
Figure~\ref{fig:amplification}(a) illustrates   a constant-amplitude surface wave propagating along an unmodulated capacitive boundary. Then, the temporal modulation of the surface reactance is suddenly  switched on.  Evidently, as it can be seen in Fig.~\ref{fig:amplification}(b, ) the surface mode is significantly amplified after modulating with a duration of $\Delta t=5T_0$, where $T_0$ is the time period at the excitation frequency.  In addition, there are higher-order harmonics generated in free space, forming a standing wave pattern along the surface, due to symmetry of the structure.


\section{Conclusions}
Here, we have presented the dispersion equation and example dispersion plots for a temporally modulated electromagnetic boundary. The results show that  time modulation induces band gaps in the two-dimensional $k$-space, which provides opportunities to control surface wave propagation. In the presentation, we will discuss in detail interesting properties of surface waves launched  inside a band gap.  
Let us note that one can repeat the above path to achieve the dispersion equation associated with  time-varying inductive boundaries. The difference is that the wave has a transverse-magnetic polarization (i.e., magnetic field is perpendicular to the propagation plane). 


\section{Acknowledgment}
The authors thank Dr. V.~Asadchy for useful discussions and valuable comments.

\bibliography{references}
\bibliographystyle{IEEEtran}

\end{document}